\UseRawInputEncoding

\documentclass[12pt]{iopart}

\usepackage{iopams}
\usepackage[version=4]{mhchem}
\usepackage{color}
\usepackage{cite}
\usepackage{graphicx}
\usepackage{ulem}
\begin{document}

    \title{The unusual distribution of spin-triplet supercurrents in disk-shaped Josephson junctions}

\author{Remko Fermin, Junxiang Yao, Kaveh Lahabi, and Jan Aarts}
\address{Huygens-Kamerlingh Onnes Laboratory \\ Leiden University \\ P.O. Box 9504, 2300RA Leiden, The Netherlands}
\ead{aarts@physics.leidenuniv.nl}
\vspace{10pt}

\begin{abstract}
The phenomenon of s-wave spin triplet Cooper pairs induced in ferromagnetic metals has been researched now for more than a decade, and its main aspects are well understood. Crucial in converting s-wave singlet pairs in the superconductor to s-wave triplets in the ferromagnet is the engineering of well-defined magnetic inhomogeneity (the 'generator') at the interface with the superconductor. Vertical layer stacks are typically used as such, where two separate thin ferromagnetic layers with homogeneous but non-collinear magnetizations, provide the inhomogeneity. Alternatively, magnetic textures, like ferromagnetic domain walls and vortices, are possible triplet generators, although they are far less studied. In this paper we review our experiments on lateral disk-shaped Josephson junctions where a ferromagnetic bottom layer provides a weak link with a vortex magnetization imposed by the shape of the disk. We present three different junction configurations, exhibiting their own generator mechanism. In the first, we utilize the non-collinearity with a second ferromagnetic layer to produce the triplet correlations. The second configuration consists of only the bottom ferromagnet and the superconducting contacts; it relies on the vortex magnetization itself to generate the spin-polarized supercurrents. In the third case we exploit an intrinsic generator by combining a conventional superconductor (NbTi) and a half-metallic ferromagnetic oxide (La$_{0.7}$Sr$_{0.3}$MnO$_3$). We find strong supercurrents in all cases. A particularly interesting finding is that the supercurrents are strongly confined at the rims of the device, independent of the generating mechanism, but directly related to their triplet nature. What causes these \textit{rim currents} remains an open question.
\end{abstract}

%
%
%
%
%

\section{Introduction}

Supercurrents of s-wave nature can be induced in ferromagnets by exploiting their anti-symmetrical pairing in the time domain. Specifically, the Pauli exclusion principle is not violated when the wave function describing the Cooper pair obtains a minus sign under exchange of time variables. Therefore, such pairing (usually called odd-frequency) can be of equal-spin and s-wave nature simultaneously. These supercurrents,  carried by spinfull triplet Cooper pairs, are therefore spin-polarized. A triplet pair in a ferromagnet is typically broken by the temperature (similar to a singlet in a normal metal), or by the spin lifetime, giving rise to a significantly longer coherence lengths than a spinless singlet pair in the same ferromagnet: the singlet Cooper pair is broken by the exchange energy, leading to a coherence length of the order of only nanometers. Due to their long coherence lengths in ferromagnets, triplet pairs form the building block for superconducting spintronics, a field that strives towards dissipationless information processing using spin\cite{Linder2015,Eschrig2015a,Robinson2021}. The possibility of robustly generating a long-range proximity effect was discussed in a seminal paper by Bergeret \textit{et al.}\cite{Bergeret2001}, who first suggested a mechanism for formation of the triplet pairs in S/F-hybrids, using a controlled local inhomogeneity of the magnetization of the F layer. Specifically, they considered a changing direction of the magnetization, as can be found in domain walls \cite{Bergeret2001a,Bergeret2005}. Generally, magnetic inhomogeneity is key to generating triplets, as elucidated for instance in Refs.\cite{Kadi2001,Eschrig2003,Houzet2007}: a spin singlet Cooper pair has to undergo spin-selective scattering to produce the spinless component of a triplet, and its quantization axis has to be rotated to become a spinfull triplet. These processes are referred to as spin mixing and spin rotation, respectively.\\

Experimentally, rather than using domain walls, magnetic non-collinearity can be engineered in a controlled way by using layer stacks with different magnetic materials and  magnetization directions. Such stacks of type F$_1$/F$_2$/F$_1$ were used early on to study the long range triplet (LRT) proximity effect. This was done by using a Ho/Co/Ho stack, which benefited from the helical nature of the ferromagnetic holmium~\cite{Robinson2010}, but also with stacks of homogeneous ferromagnets such as weakly ferromagnetic PdNi or strongly ferromagnetic Ni for the F$_1$ layer~\cite{Khaire2010,Khasawneh2011}. Most of the subsequent work relied on such stacks as triplet generator, although LRT supercurrents were also reported in singular ferromagnets such as Co~\cite{wang2010} or the Heusler alloy \ce{Cu2MnAl}~\cite{sprungmann2010}.\\

Special mentioning deserves the case of the ferromagnetic metallic oxides \ce{CrO2} and La$_{0.7}$Sr$_{0.3}$MnO$_3$ (LSMO), that are both half metals, meaning they are fully spin-polarized. In that case, the range of the proximity effect can be expected to be particularly long. An early report of a \ce{CrO2} film with NbTi contacts grown on a \ce{TiO2} substrate, but without obvious triplet generator, reported supercurrents flowing over a distance of almost 0.5~{\textmu}m \cite{Keizer2006}. Similar results were found for \ce{CrO2} films grown on \ce{Al2O3} substrates~\cite{Anwar2010}. Later work showed that a generator stack could be fruitfully used to yield large spin valve effects~\cite{Singh2015,Voltan2016} and large supercurrent densities.~\cite{Anwar2012,Singh2016} Triplet generation in \ce{CrO2} without such a stack is currently believed to be due to intrinsic strain-induced magnetic inhomogeneity (for films on \ce{TiO2}) or grain boundary disorder (for films on \ce{Al2O3}).~\cite{Anwar2011}  For LSMO, the picture is intriguing: supercurrents were reported in hybrid structures with the high-T$_c$ superconductor \ce{YBa2Cu3O7}, both for vertically stacked structures~\cite{Visani2012}, and recently for a lateral junction~\cite{Sanchez2021}. Since there are no established sources of magnetic inhomogeneities in these structures, the mechanism for triplet generation is not yet clear.\\

 \begin{figure}[h!]
 \centerline{$
 \begin{array}{c}
  \includegraphics[width=1\linewidth]{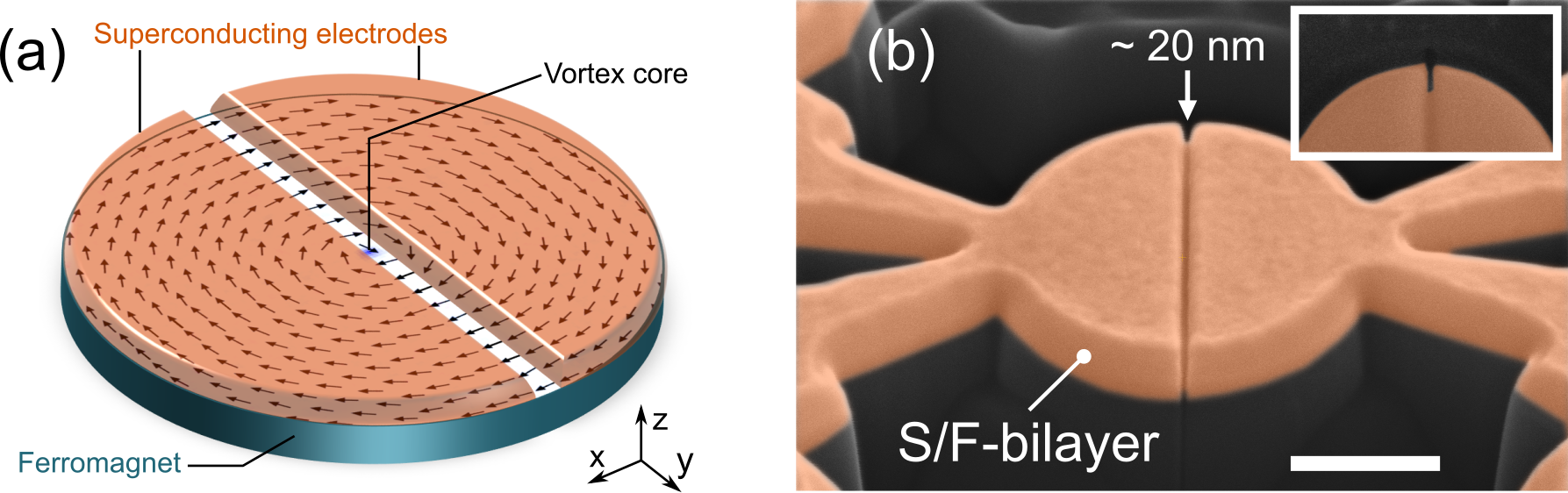}
 \end{array}$}
 \caption{\textbf{(a)} Schematic of the disk-shaped Josephson junctions. The superconducting electrodes are separated by a trench, forming a ferromagnetic weak link. The pattern on the ferromagnetic layer corresponds to micromagnetic simulations of a micron-size disk. The arrows correspond to the in-plane magnetization, while the out-of-plane component is represented by color, which only appears at the vortex core (blue region; less than 5 nm in diameter). \textbf{(b)} False colored scanning electron micrograph of a structured bilayer. The 20 nm gap indicates the weak link at the bottom of the trench. The scale bar is equivalent to 400 nm. The inset shows a zoom of a notch formed by the geometry dependent milling rate. The image is from a different disk, fabricated in the same way. Figure adapted after one in Ref.~\cite{Co_disk_paper}.}\label{disk}
 \end{figure}

%
More recently, alternative methods of generating LRT currents were discussed. Specifically, spin textures, such as domain walls~\cite{Bergeret2001a,Fominov2007,Kalcheim2011,Aikebaier2019} or magnetic vortices~\cite{Kalenkov2011,Silaev2009} were predicted as generators. Besides, spin mixing can also be achieved by spin-orbit coupling (SOC), leading to various theoretical considerations~\cite{Niu2012,Bergeret2013_New,Bergeret2014a,Alidoust2015,jacobsen2015,Bujnowski2019,Eskilt2019,Silaev2020}. Although this spawned various experiments on Josephson junctions with heavy metal interlayers~\cite{Satchell2018,Jeon2018,Satchell2019}, experiments on junctions with controlled spin textures remain scarce.\\

In this paper, we review the research we have carried out on mesoscopic planar Josephson junctions, featuring a well-defined spin texture induced by the disk shape of the device. Here the disk geometry brings two advantages with respect to classic stacked junctions: stray fields are (almost) absent and the disk geometry allows us to directly study the role of a vortex magnetization on the generation of triplet supercurrents. The concept is shown in Figure \ref{disk}a. The device consists of a disk-shaped bilayer or layer stack. Here the bottom ferromagnetic layer exhibits a ferromagnetic vortex magnetization pattern, whereas the top layer is split into two superconducting electrodes. This is done by locally removing the superconductor in the middle of the disk. Since we exclusively remove the superconductor, the current path is forced through the ferromagnet below, which creates an S/F junction.\\

The paper is organized as follows. First we present our basic study object, the planar disk-shaped planar Josephson junction; we describe the device fabrication, and we show the results on singlet junctions made from a superconductor and a normal (N) metal. By analyzing the dependence of the critical current on an out-of-plane magnetic field $I_c(B_{\perp})$, we demonstrate that in such S/N junctions the current distribution is fully homogeneous. In contrast, when the bottom layer is replaced by a ferromagnet, we find the triplets supercurrents to flow in highly localized channels at the rims of the device. We review the similarities and differences between three types of ferromagnetic junctions. The first consists of a stack Nb/Ni/Cu/Co, where the design philosophy was that of magnetic non-collinearity of stacked junctions. The second type is based on a Nb/Co bilayer, and makes full and only use of the spin texture to generate LRT currents. In the third we use LSMO as the ferromagnet, which is a special case due to its fully spin-polarized charge carriers.

\section{Planar disk junctions}

\subsection{Fabication of disk-shaped junctions}

Central to the fabrication of the junctions is the focused Ga$^{+}$-ion beam (FIB) milling procedure. After lithographically defining a four-probe geometry and sputter depositing the bilayer or layer stack, we use FIB milling to define the shape of the junction. By applying an ultra-low beam current of 1.5 pA, the weak link is formed by a line cut in the superconducting layer along the diameter of the device. Generally, forming this trench separates the superconducting electrodes by a roughly $20$ nm weak link, allowing for Josephson coupling. In Figure \ref{disk}b we show a false colored scanning electron micrograph of a typical disk-junction. The cut becomes slightly deeper at the edge of the structure due to a geometrical dependence of the milling rate, leading to a small notch of order 50~nm on the sides of the disk, making the diameter of the weak link slightly smaller than the disk diameter. The notches, on the side of the disk, can be seen in Figure \ref{disk}b.

\subsection{S/N disk junctions: MoGe/Ag}\label{normal_section}

\begin{figure}[t]
 \centerline{$
 \begin{array}{c}
  \includegraphics[width=1\linewidth]{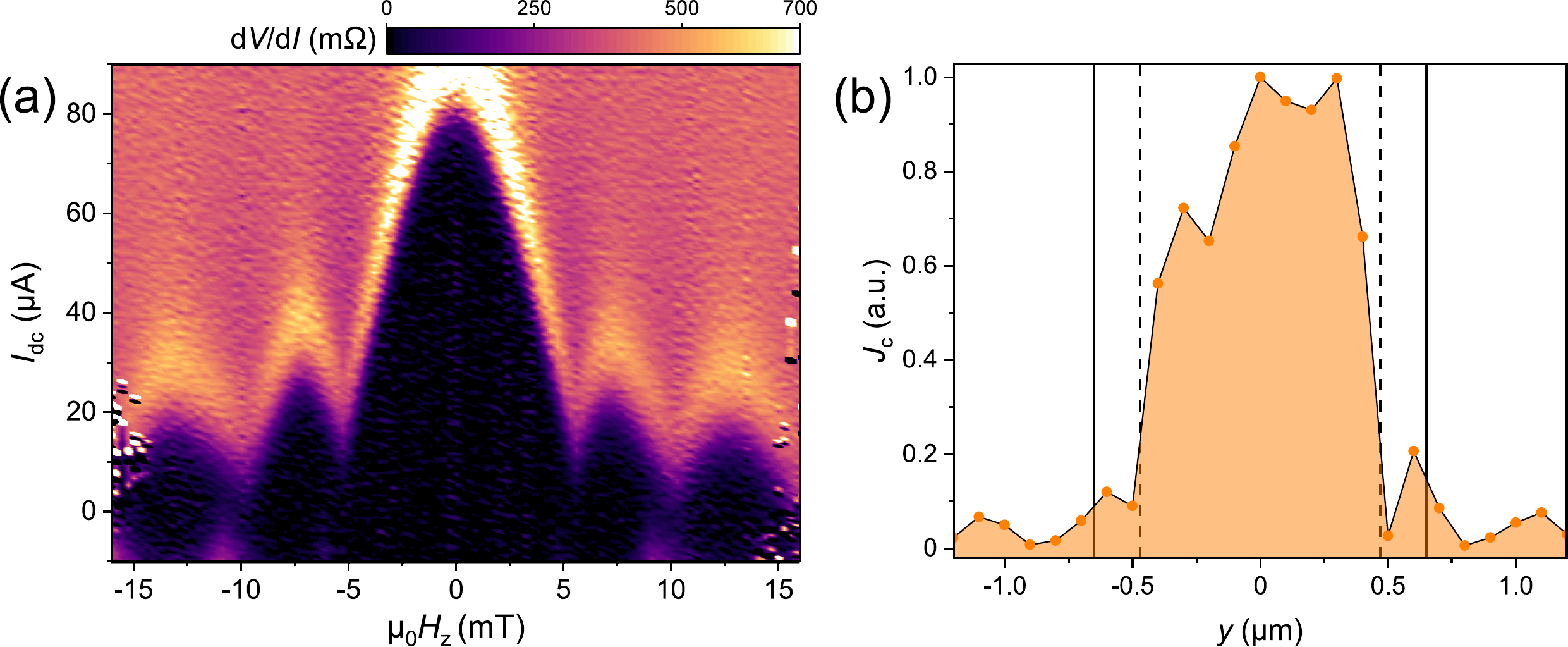}
 \end{array}$}
 \caption{\textbf{(a)} $I_c(B_{\perp})$ measurement of a disk-shaped S/N Josephson junction as a d$V/$d$I$ color map. As expected for an S$-$N$-$S junction, we observe a typical Fraunhofer interference pattern. in \textbf{(b)} we plot the critical current distribution, obtained from Fourier analysis of the data shown in (a). We indicate the sides of the electrodes by solid vertical reference lines; the dashed lines indicate the sides of the actual weak link. Figure adapted after Ref.~\cite{Beyond_the_effective_length}.}\label{sqi-sn}
 \end{figure}

To confirm that we can reliably fabricate disk-shaped Josephson junctions and rule out any secondary effect induced by the shape of the superconducting electrodes, we fabricated junctions with a normal metal weak link from a MoGe (55 nm)/Ag (20 nm) bilayer. We use $I_c(B_{\perp})$ measurements to confirm the uniformity of the critical current in these S/N devices. We determine $I_c(B_{\perp})$ of the device, with $\mu_0H_z = B_{\perp}$ an out-of-plane magnetic field (perpendicular to the current flow), by means of current ($I$) versus voltage ($V$) measurements. A color plot of the resulting $I_c(B_{\perp})$ pattern is shown in Figure \ref{sqi-sn}a. For the normal metal weak link junctions, we observe a typical Fraunhofer interference pattern corresponding to a single Josephson junction with a uniform critical current distribution.\\

To analyze the $I_c(B_{\perp})$ patterns in more detail, we perform a Fourier analysis to extract the critical current density distribution using a technique originally proposed by Dynes and Fulton~\cite{Dynes1971}. In order to do so, we first extract $I_c$ from the $I_c(B_{\perp})$ pattern by a voltage threshold criterion. Next we use a complex inverse Fourier transform to extract the critical current density distribution from $I_c(B_{\perp})$. The Fourier analysis is based on knowledge of the shielding currents in the electrodes. For junctions between macroscopic superconducting electrodes, these are given by the Meissner effect. As a consequence, the effective length of such junctions is given by $d + 2 \lambda_{\text{L}}$, where $d$ is the length of the junction (i.e., distance between the superconducting electrodes) and $\lambda_{\text{L}}$ the London penetration depth. However, since our junctions are in the thin film limit (i.e., the superconducting layer thickness is smaller than $\lambda_{\text{L}}$) and our junctions are laterally constricted in a disk geometry, the shielding currents are given by a non-local electrodynamic relation, which alters the effective junction length in a significant way. We recently developed a method to evaluate the shielding currents using simulations of the gauge-invariant phase gradient in our junctions. A detailed discussion of these simulations and the technical details of the Fourier analysis can be found elsewhere~\cite{Beyond_the_effective_length}.\\

Figure \ref{sqi-sn}b shows the critical current density distribution corresponding to the $I_c(B_{\perp})$ pattern of Figure \ref{sqi-sn}a. Here we indicate the width of the electrodes by solid vertical reference lines. By dashed lines we indicate the width of the weak link itself, which is slightly smaller due to the aforementioned notches. As expected for a single junction, we observe a uniform distribution of critical current throughout the weak link. This means we can reliably fabricate disk-shaped Josephson junctions using our FIB fabrication technique.

 \begin{figure}[b!]
 \centerline{$
 \begin{array}{c}
  \includegraphics[width=0.75\linewidth]{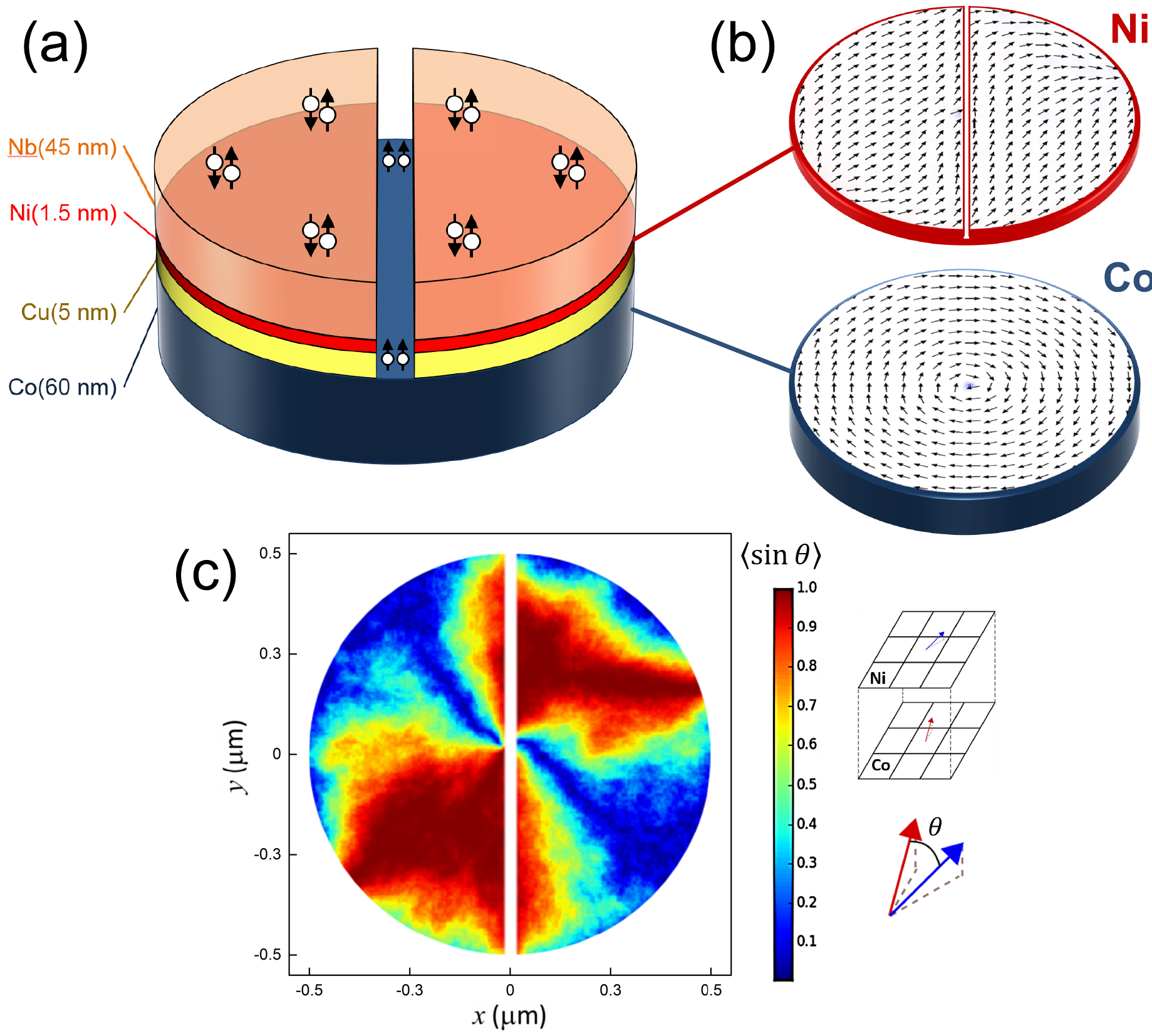}
 \end{array}$}
 \caption{(a) Schematic of the Nb/Ni/Cu/Co device layout, showing the different layers and their thickness. The Nb layer contains singlet Cooper pairs, the current in the Co is carried by triplets. (b) The magnetic texture in the Co and Ni layers, calculated by 3D simulations using the OOMMF package. Note that the magnetic moments in Ni tend to align with the gap and therefore perpendicular to the magnetic moments in the Co. (c) The magnetic non-collinearity profile in terms of the calculated angle $\theta$ between the magnetization vectors in the Co and the Ni. The red regions are areas of large non-collinearity. Figure adapted after one in Ref.~\cite{Lahabi2017a}.}\label{disk-sf}
 \end{figure}

\section{Superconductor-ferromagnet junctions}
\subsection{Magnetic non-collinearity:  Nb/Ni/Cu/Co disk junctions} 

The first type of ferromagnetic junction consists of a layer stack of Co (60 nm)/Cu (5 nm)/Ni (1.5 nm)/Nb (45 nm), where we utilize the non-collinearity of the two magnetic layers to generate triplet Cooper pairs. We reported results on this device in Ref.~\cite{Lahabi2017a} but re-analyzed $I_c(B_{\perp})$ according to our more recent insights~\cite{Beyond_the_effective_length}. Here we form the ~20 nm Co barrier by a trench that cuts through the layers above (see Fig.~\ref{disk-sf}a). Micromagnetic simulations confirmed that the bottom (Co) layer has a vortex magnetization, while shape anisotropy would force the Ni magnetization along the edges of the trench, as schematically shown in Fig.~\ref{disk-sf}b. Note that the Cu layer decouples the two magnetic layers. The calculated amount of non-collinearity in terms of the angular difference between the local magnetization directions in the Co and the Ni is shown in Fig.~\ref{disk-sf}c. The values are high along the trench, except for a small region at the center, where the out-of-plane field from the vortex core (about 20 nm in diameter) locally couples to the Ni magnetization. \\

 \begin{figure}[t!]
 \centerline{$
 \begin{array}{c}
  \includegraphics[width=1\linewidth]{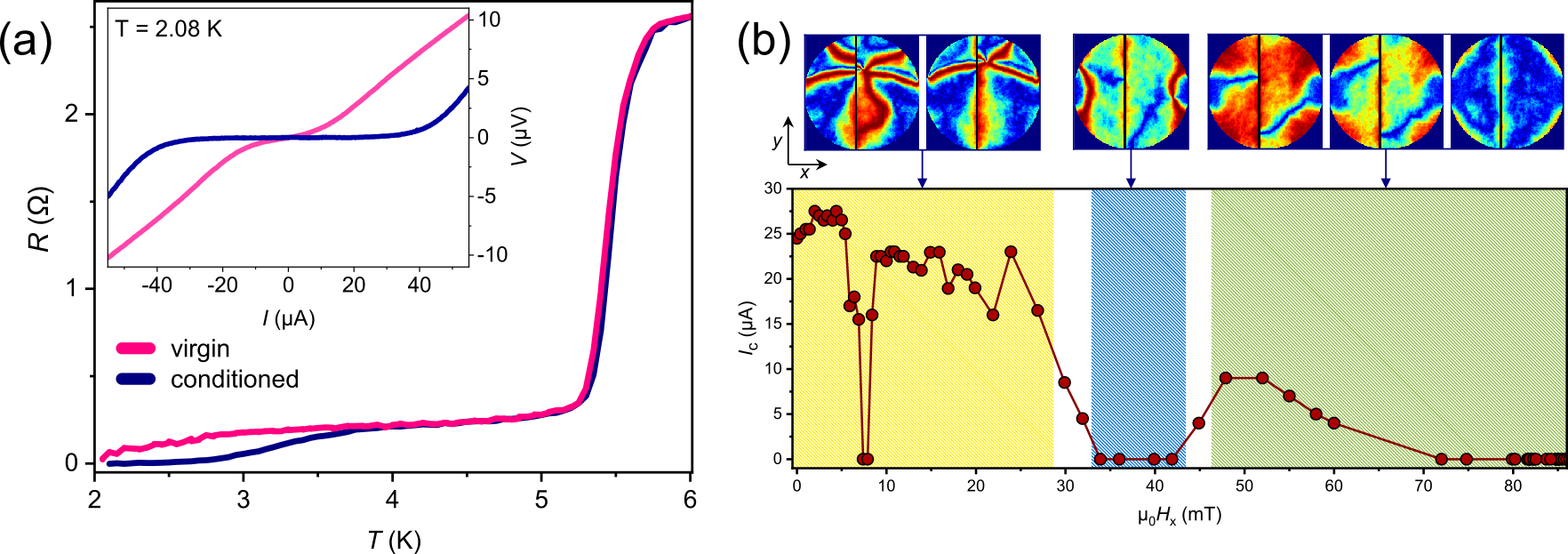}
 \end{array}$}
 \caption{(a) The dependence of the resistance $R$ of the Nb/Ni/Cu/Co device on temperature $T$ for the virgin (pink) state and the conditioned (blue) state, measured using a current of 10~{\textmu}A. The inset show two $I$-$V$ traces (current-voltage), taken at 2.08~K for the virgin (pink) and the conditions (blue) case. (b) Critical current variation and simulations of the magnetic non-collinearity as function of an in-plane field $B_x$ directed perpendicular to the trench. Note the motion of the vortex core is directed along the trench towards the side of the disk. The vortex state in Co disappears around 35~mT and the Ni layer becomes magnetized antiparallel to the Co. Above 45~mT, the field starts to align the Ni magnetization against the stray fields from the Co layer, leading again to non-collinearity. The alignment is complete above 60~mT, and superconductivity is suppressed. Figure adapted after one in Ref.~\cite{Lahabi2017a}.}\label{RT-Ic}
 \end{figure}
%

Typical to the Nb/Ni/Cu/Co devices is that magnetic conditioning of the nickel layer is required to achieve the non-collinearity, as is also evidenced by our transport experiments. Fig.~\ref{RT-Ic}a compares the resistance $R$ versus temperature $T$ of the device and IV measurements taken before and after conditioning the device with an out-of-plane field of 2.5~T. In the virgin state, hardly any supercurrent is observed below the initial superconducting transition (5.5~K), but after conditioning, $R$ goes fully superconducting below 3~K and the $I$-$V$ characteristics (inset Fig.~\ref{RT-Ic}a) show a well-defined zero-voltage state with an enhanced $I_c$. Next we examined the precise influence of the spin texture in the Co layer by applying an in-plane field $B_x$ along the x-direction of the device, which is perpendicular to the trench, thereby moving the vortex core along the trench. The results, complemented with non-collinearity maps from the micromagnetic simulations, are given in Fig.~\ref{RT-Ic}b. The first thing to note, is a deep dip in $I_c$ around 8~mT. This is a surprisingly robust feature, which cannot be explained by the magnetic non-collinearity, also occurs in Co disks without the Ni layer. As we discuss in the next section, this dip appears to signal a 0 to $\pi$ phase shift in the junction, which results from the asymmetric spin texture of a displaced vortex. Next we observe $I_c$ to go to zero around 34~mT. At this field the vortex core leaves the disk, and stray fields that emerge from the Co-layer align the magnetization in the Ni layer antiparallel. A further increase of the field starts to align the Ni and Co magnetizations, but in the process non-collinearity occurs again, and superconductivity reenters. Only around 70~mT, when both magnetizations are aligned, superconductivity disappears. \\

 \begin{figure}[t]
 \centerline{$
 \begin{array}{c}
  \includegraphics[width=1\linewidth]{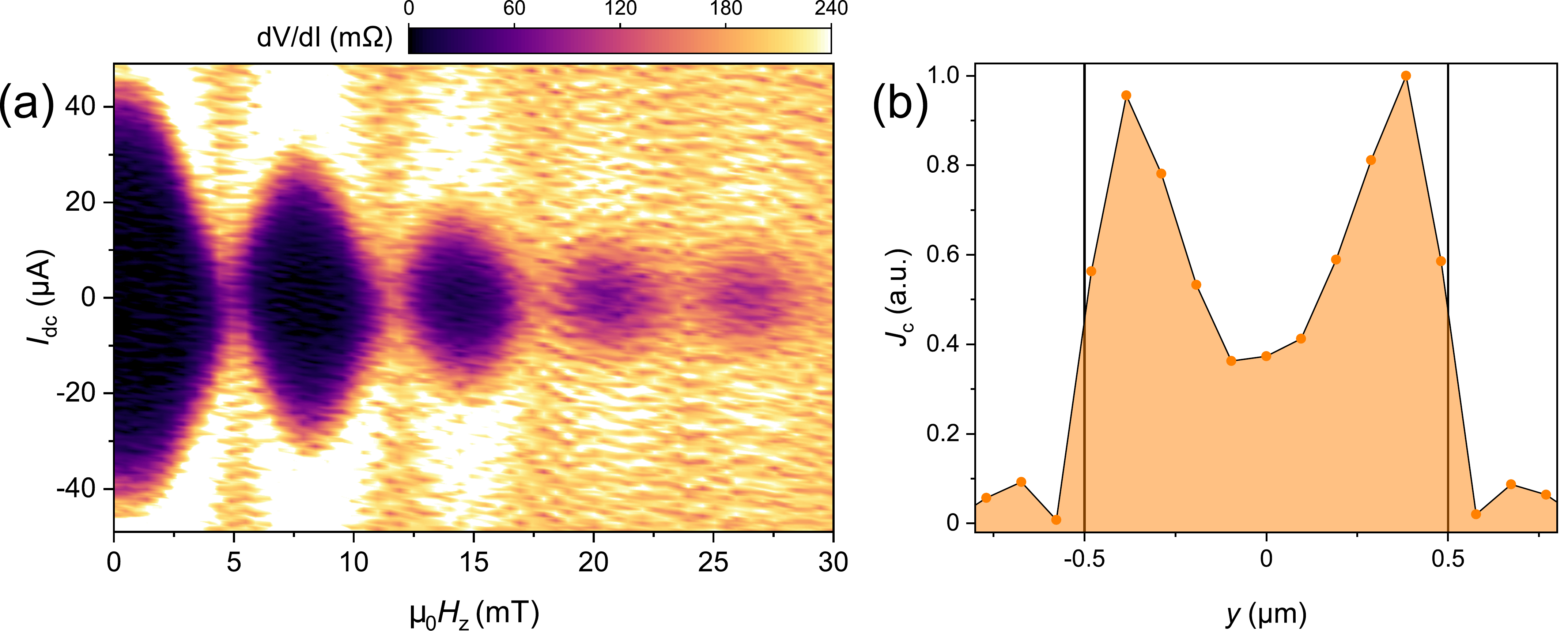}
 \end{array}$}
 \caption{(a) The critical current $I_c$ of the Nb/Ni/Cu/Co device as function of an out-of-plane magnetic field $\mu_0H_z$ taken at 2.8~K. (b) The current density profile constructed from the Fourier analysis. The sides of the electrodes are shown by the solid lines. The critical supercurrent density is predominantly distributed on the sides of the junction.}\label{sqi-nbnico} 
 \end{figure}
%
 
Next we consider the critical current distribution in the junction by measuring $I_c(B_{\perp})$ (at zero in-plane magnetic field). The result, given in Fig.~\ref{sqi-nbnico}a is surprising, and qualitatively different from the S/N case. The first lobe in the pattern is wider than the next ones, but less than expected for a Fraunhofer pattern. It rather tends to a two-channel interference pattern, where all lobes have equal width. The Fourier analysis of the pattern confirms the presence of two parallel channels. The supercurrent distribution profile presented in Fig.~\ref{sqi-nbnico}b, shows that the current is inhomogeneous, with clear peaks at the rims of the disk. Apparently, the current flows primarily around the rim, although appreciable transport still takes place via the center of the disk. Note that we re-analyzed the data with the nonlocal electrodynamics approach, removing the ambiguity about the actual diameter of the sample~\cite{Lahabi2017a}.\\

Reiterating the most salient points, this device shows long range triplet supercurrents through the Co weak link. The triplet supercurrents are generated by the amount of magnetization non-collinearity of the two F-layers and the experimental response of the device can be almost completely determined by this magnetic non-collinearity. However, the critical current density at the rims of the device is substantially larger than in the center. Besides, there is a robust dip in the critical current as function of in-plane magnetic field, which cannot be explained by the magnetic non-collinearity of the Co and Ni. These two observations will come back in the next two paragraphs.

\subsection{LRT generation using a ferromagnetic vortex: Nb/Co disk junctions}

As discussed in the previous section, we can capture much of the characteristics of the Nb/Ni/Cu/Co junctions by the non-collinearity of the Co and Ni layers. However, there remain a couple of open questions. One is: why a portion of the supercurrent flows along the sides of the junction; another is why is there a sudden decrease of $I_c$ at an in-plane magnetic field of around 10 mT. Besides these questions raised by our experiments, there exist multiple theoretical proposals that discuss the generation of long-range triplet correlations using the vortex magnetization of the weak link itself~\cite{Silaev2009,Kalenkov2011}. This motivated the study of Nb/Co devices that lack the Ni and Cu layer~\cite{Co_disk_paper}. Any superconducting correlations in these devices must be solely generated by the spin texture of the ferromagnetic layer.\\

Indeed, we find that the Nb/Co devices a show long-range proximity effect, with similar critical current values as those measured in the Nb/Ni/Cu/Co junctions. We have verified that the naturally present vortex spin texture is responsible for the generation of long-range triplet correlations by showing that the critical current of the Nb/Co junctions vanishes when the magnetization is uniform (either by magnetizing the Co-layer or by various control experiments). Besides, the Nb/Co junctions show no difference between the virgin or field conditioned states, as is the case with the Nb/Ni/Cu/Co devices. This difference is caused by the fact that, unlike the 1.5~nm Ni layer in the electrodes, the Co disk has a highly stable vortex ground state, which does not require magnetic conditioning.\\

 \begin{figure}[t!]
 \centerline{$
 \begin{array}{c}
  \includegraphics[width=1\linewidth]{Figure6_V2.pdf}
 \end{array}$}
 \caption{\textbf{(a)} and \textbf{(b)} Show $I_c(B_{\perp})$ patterns obtained on two Nb/Co disk junctions with different diameters. The disk diameters in (a) and (b) are 1.62~{\textmu}m and 1.05~{\textmu}m, respectively. The period of the oscillations scales inversely with the junction area. In both cases, the junctions show a clear two-channel interference pattern. (c) and (d) show the critical current density distributions obtained by the Fourier analysis of the patterns in (a) and (b) respectively. The vertical lines indicate the boundaries of the electrodes of the device, whereas the dashed lines indicate the sides of the actual weak link. Figure adapted after one in Ref.~\cite{Co_disk_paper}~\protect\footnotemark.}\label{sqi-nbco}
 \end{figure}
 
We investigated the critical current distributions in the Nb/Co devices by $I_c(B_{\perp})$ measurements; the results are plotted in Figure \ref{sqi-nbco}, along with the critical current distributions evaluated using inverse Fourier transform. Specifically, we show the interference patterns for two junctions with different diameters (1.62 and 1.05~{\textmu}m) in Figure \ref{sqi-nbco}a and b. Note that the period of the oscillations scales inversely with the area of the junction, which is determined by the radius of the disk. This implies that we consistently find the supercurrent to be highly localized in 70 nm wide channels at the rims of the sample, regardless of the sample area. We therefore call these currents \textit{rim currents}. As a control experiment, we also prepared a disk junction with a relative shallow trench. This provides a non-magnetic weak link for singlet correlations. In that case we observed a typical Fraunhofer-like interference pattern with a two times wider central lobe (not shown here). In reference \cite{Co_disk_paper}, we suggested a model to explain the appearance of the rim currents in terms of a spin accumulation at the edges due to an effective orbit coupling generated by the vortex magnetization. We will further detail this model in the discussion in section \ref{discussion}.\\

Comparing the critical current distributions in Figure \ref{sqi-nbco}b and d to those obtained on the Nb/Ni/Cu/Co devices (Figure \ref{sqi-nbnico}b), we observe that both devices share a tendency for supercurrent to flow along the rims of the device. The difference, however, is the relative distribution of rim currents: the Nb/Co junctions show a far stronger concentration of critical current on the rims than the Nb/Ni/Cu/Co devices. Finally, note that the two-channel behavior is absent in all our control samples that are either singlet dominated (e.g., those presented in section \ref{normal_section}) or lack any spin texture (by fully magnetizing the Nb/Co devices).\\

 \begin{figure}[t]
 \centerline{$
 \begin{array}{c}
  \includegraphics[width=1\linewidth]{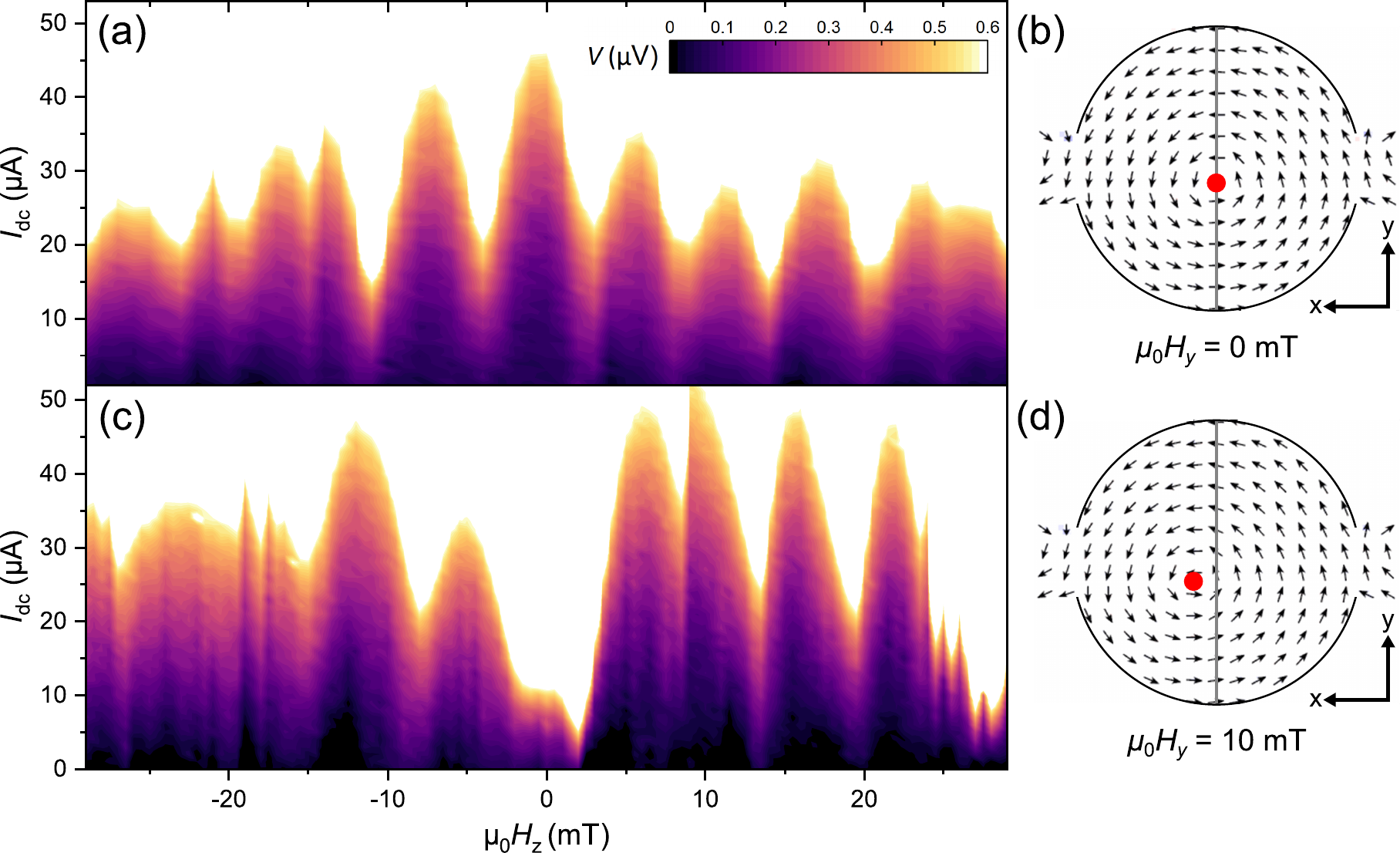}
 \end{array}$}
 \caption{$I_c(B_{\perp})$ patterns (left column) obtained on Nb/Co junctions, measured at different in-plane fields and the corresponding simulated spin textures (right column). In the simulations, the gray line represents the position of the weak link, and the red dot indicates the location of the vortex core. (a) and (b) are obtained at zero in-plane field: the vortex is at the center of the disk, and a two channel pattern is observed, i.e., lobes of equal width and slow decay of peak height. (c) and (d) Applying $\mu_0 H_\text{y}$ = 10 mT breaks the axial symmetry of the vortex magnetization, by effectively displacing it away from the junction. This results in the suppression of the middle peak in the interference pattern, characteristic of a $0-\pi$ SQUID. Figure adapted after one in Ref.~\cite{Co_disk_paper}~\footnotemark[1].}\label{nbco-inplane}
 \end{figure}
 
\footnotetext{Reference \cite{Co_disk_paper} can be found at: https://pubs.acs.org/doi/10.1021/acs.nanolett.1c04051. Further permission related to the reuse of these figures should be directed to the ACS.}

To further investigate the interplay between spin texture and the generation of long-range triplet supercurrents, we modify the vortex magnetization pattern using in-plane magnetic fields and study the result on the critical current distribution. In this case, we apply a magnetic field along the trench direction (the y-direction) and effectively move the vortex away from the junction (towards the contacts). This displacement is linear for small in-plane fields. While maintaining a static in-plane field, we can sweep the out-of-plane magnetic field by the use of a vector magnet. The $I_c(B_{\perp})$ patterns for a Nb/Co disk-device at zero in-plane magnetic field and 10 mT in-plane field are compared in Figure \ref{nbco-inplane}. At zero in-plane field, we observe a clear two-channel interference pattern, characterized by lobes of equal width and a maximum of $I_\text{c}$ at zero out-of-plane field. Similarly, at $\mu_0 H_\text{y}$ = 10 mT we observe a pattern featuring equal-width lobes, but now with a strong suppression of $I_\text{c}$ at zero out-of-plane field. The resulting $I_c(B_{\perp})$ pattern therefore bears resemblance of a $0-\pi$ SQUID. Such a $0$ to $\pi$ transition is also observed in our Nb/Ni/Cu/Co devices, where a $\sim$10~mT in-plane field perpendicular to the trench results in a sharp drop in critical current (see Fig.~\ref{RT-Ic}b). Note that in the Nb/Co case the in-plane field is along the trench and in the Nb/Ni/Cu/Co case the in-plane field is perpendicular to the trench. Still, we conclude that the $0$ to $\pi$ transition is a universal feature of the vortex magnetization of the Co-layer and the relative displacement of the vortex core from the center of the disk.

\subsection{Half metallic ferromagnet: NbTi/LSMO disk junctions}

The third system we studied is similar to the Nb/Co system: it does not rely on the relative orientation of multiple magnetic layers but consists out of a single ferromagnet. However, instead of a $3d$ ferromagnetic transition metal, we use the halfmetallic ferromagnetic oxide La$_{0.7}$Sr$_{0.3}$MnO$_3$ (LSMO). The halfmetallic nature of LSMO means that only one type of triplet pair can exist, either spin-up or spin-down. We have also replaced the Nb for NbTi superconducting electrodes.\\

 \begin{figure}[h!]
 \centerline{$
 \begin{array}{c}
  \includegraphics[width=1\linewidth]{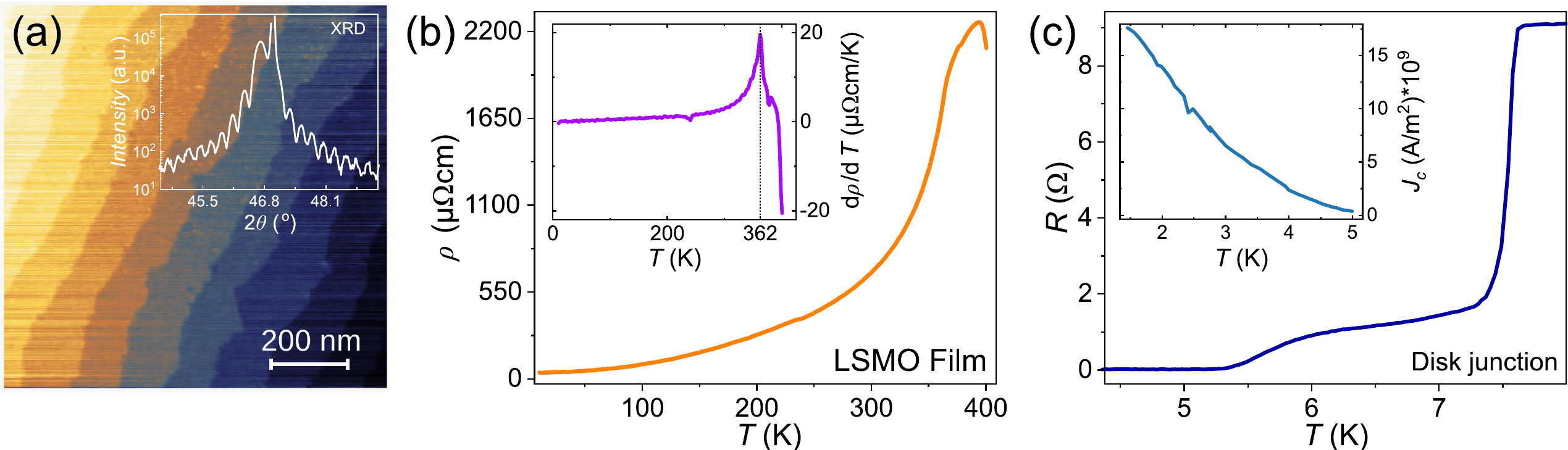}
 \end{array}$}
 \caption{(a) Atomic Force Microscopy image of a 40 nm thick LSMO film grown on an LSAT substrate. Terraces can be seen. The inset shows an xray difraction measurement (XRD). Intensity oscillations around the main Bragg reflection at 47$^{\circ}$ signals a high flatness of the film. (b) Resistivity $\rho$ versus temperature $T$ of a 40 nm thick LSMO film. The Curie temperature can be read off from the peak of the derivative shown in the inset (362 $^{\circ}$C). (c) Resistance $R$ versus $T$ of a NbTi/LSMO disk junction. The inset shows the temperature dependence of the critical current density $J_c$.}\label{lsmo}
 \end{figure}

Using the method described above, a disk-shaped NbTi/LSMO junction was made, with similar dimensions as the Co ones. We first grow a thin (40~nm) LSMO film on an (La$_{0.18}$Sr$_{0.82}$)(Al$_{0.59}$Ta$_{0.41}$)O$_3$ (LSAT) substrate by off-axis sputtering (growth pressure 0.7 mbar in an Ar:O (3:2) atmosphere; background pressure about 10$^{-7}$~mbar) at a substrate temperature of 700~$^o$C. The substrate is chosen to minimize the mismatch between film and substrate. The LSMO is of high quality, as can be seen from the x-ray diffraction and atomic force microscopy data in Fig.~\ref{lsmo}a. Fig.~\ref{lsmo}b shows the R(T) data of a typical film. The ferromagnetic ordering temperature (the Curie temperature) is at about 360~K, and the resistivity drops by almost two orders of magnitude upon decreasing the temperature, with a value of 40~\textmu$\Omega$cm at 10~K. In the same system, we sputter-deposit a NbTi film of about 60~nm with a superconducting transition temperature of about 7.5~K.\\

Fig.~\ref{lsmo}c shows R(T) of the device, with the typical signature of a proximity effect: first the contacts go superconducting, then R(T) shows a plateau where the weak link is still in the normal state, followed by a second transition and zero resistance reached just below 5.5~K.The inset shows the critical current density $J_c$ as function of temperature. $J_c$ steadily increases, reaching 1.8~$\times 10^{10}$~A/m$^2$ at the lowest temperature of 1.5~K. This is an even higher value than found in the disk junctions discussed before. For example, the Nb/Co disk of the previous section, we estimate the current density in the channels to be around 8~$\times 10^{9}$~A/m$^2$. High values are consistent with the earlier works on long-range triplet proximity in half-metals (e.g., \ce{CrO2} junctions and spin valves). As far as we are aware, there are no previous reports on the proximity effect in LSMO or similar metallic oxides with a conventional superconductor.\\

\begin{figure}[t]
 \centerline{$
 \begin{array}{c}
  \includegraphics[width=1\linewidth]{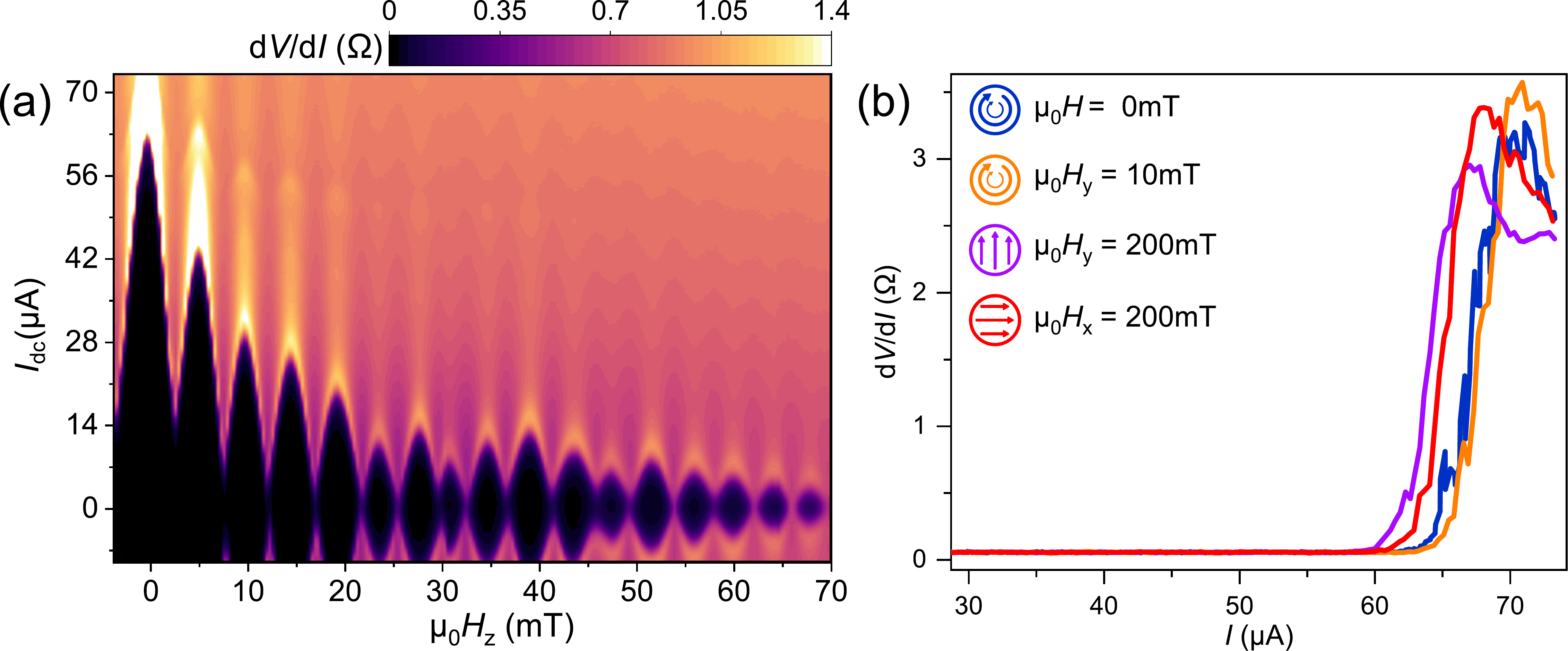}
 \end{array}$}
 \caption{(a) $I_c(B_{\perp})$ pattern recorded at 4.1~K on a disk-shaped NbTi/LSMO junction. (b) IV-characteristics under the application of constant in-plane fields parallel or perpendicular to the junction of the device. The illustrations in the legend demonstrate the corresponding magnetization states. Note that the critical current is hardly suppressed although no ferromagnetic vortex is present in the LSMO layer.}\label{sqi-lsmo}
 \end{figure}

The $I_c(B_{\perp})$ pattern of the device, taken at 4.1~K, is shown in Fig.\ref{sqi-lsmo}(a). We again observe a clear two channel interference pattern, similar to what is seen in the Nb/Co system. The current distribution as calculated by Fourier analysis (not shown) confirms that also in the NbTi/LSMO disk junctions the supercurrent distribution strongly peaks at the rims of the device. However, the behavior upon applying an in-plane field is very different, as seen Fig.\ref{sqi-lsmo}(b). Under the application of in-plane fields (either parallel or perpendicular to the trench), the critical current is quite robust: there is only a negligible decrease of critical current upon increasing the field to 200~mT. Note that 200~mT is sufficient to remove any spin texture in the disk-shaped NbTi/LSMO junction. Besides, the $0-\pi$ transition observed in the Co-based systems when moving the vortex core is absent. However, the aforementioned rim currents persist regardless of the magnetization state. Therefore, it appears that the triplet supercurrents arise from an intrinsic mechanism that is not changed by even strong in-plane fields.

%
 %

%
\section{Discussion and Conclusions}\label{discussion}

In this section we will review the similarities and differences of the three different magnetic disk junctions. The Nb/Ni/Cu/Co system is, at first sight, mostly understood. The non-collinearity of the Ni/Cu/Co stack appears to act as the generator, and the supercurrent disappears when an in-plane field has homogeneized the magnetizations. There is also an effect of the vortex magnetization of the Co-disk, however, since changing that spin texture by moving the vortex core with a small in-plane field leads to a 0-$\pi$ transition. However, the non-collinearity cannot explain why the current distribution would peak at the rims of the disk.\\

That brings us to the Nb/Co system. Here, the current distribution peaks more sharply at the rims of the device. In Ref.~\cite{Co_disk_paper} we suggested a model to explain the rim currents. Within the framework of the linearized Usadel equations, the vortex spin texture is equivalent to an effective spin-orbit coupling (SOC). Such an effective SOC would give rise to a spin current directed along the local magnetization direction\footnote{The spin current is carried by the spinless component of a triplet condensate naturally present at the interface between a ferromagnet and a superconductor.}. When this spin current encounters a vacuum boundary (i.e., the rims of the disk), the spin current accumulates. Since the spin current across that vacuum boundary has to be zero, a source of long range triplets emerges to generate an opposing spin current. Looking back at the Nb/Ni/Cu/Co system, that appears to show a combination of the conventional mechanism of magnetic non-collinearity coming from two F layers, {\it and} the mechanism we just described for the Nb/Co disks. The end result is a current distribution peaked at the rim, but finite in the center.\\

Shifting attention to the third ferromagnetic system: the observation that LSMO can be proximized with a conventional superconductor is an important result in its own regard. It might appear that all the ingredients from the Nb/Co case are also at play in the NbTi/LSMO system, since the supercurrent peaks at the rim. That cannot explain, however, the fact that these supercurrents are fully insensitive to in-plane fields, both directed along the trench or perpendicular to the trench. Once the magnetization becomes homogeneous, which we certainly expect for a 200~mT field, the vortex magnetization along with its effective SOC disappears, but the supercurrent is almost unchanged. Apparently, the LRT correlations in the NbTi/LSMO devices do not result from the vortex magnetization of the disk. Rather, the observations point to a magnetically disordered layer at the NbTi/LSMO interface, which can have various origins but is insensitive to the in-plane fields. This picture is attractive in the sense that it could equally well explain the generation of LRT correlations in the reports mentioned earlier~\cite{Visani2012,Sanchez2021}. What is not easy to understand from this picture is why the current peaks at the rim, if that is not due to an effective SOC. This requires better understanding of the magnetic landscape in the LSMO disk. \\

In summary, we have researched the behavior of lateral disk-type Josephson junctions for several different types of superconductor-ferromagnet configurations. Our most important finding is that triplet currents are generated in such junctions. The spin texture of the ferromagnetic disk plays an important role here, as evidenced by the strong dependence of the critical currents and their distribution over the junctions on the micromagnetic texture of the devices. In the Nb/Ni/Cu/Co devices, the generator is provided by the non-collinearity of the magnetic layers. However, by studying the Nb/Co devices, we found that the ferromagnetic vortex itself can generate triplet correlations as well. Looking back, the behavior of the Nb/Ni/Cu/Co devices can be completely described by a combination of the vortex-related effects observed in the Nb/Co case and magnetic non-collinearity. What is still less clear is why the critical supercurrent density peaks at the rim of the device: although our model based on an effective SOC captures the behavior well in the Nb/Co case, it cannot explain the appearance of rim currents in the LSMO-based devices. In these devices, the triplet correlations seem to be generated by an intrinsic mechanism at the NbTi/LSMO interface. What causes the unusual concentration of the triplet currents at the rims of the device remains an open question. We want to end by itemizing some of the most salient observations to facilitate further understanding: \\

\begin{itemize}
\item[i]
Rim currents only emerge when transport is carried out by spin-polarized triplet Cooper pairs. The two-channel interference pattern corresponding to rim currents, appear in all the disk junctions with a strong ferromagnetic barrier (Co or LSMO), where only long-range triplet correlations can survive. In contrast, the disk junctions with a non-magnetic barrier (singlet transport) all show a standard Fraunhofer diffraction pattern corresponding to a single transport channel. 

\item[ii]
The rim currents seem to emerge independently of the mechanism behind the generation of long-range triplet correlations. The different types of disk systems discussed above suggest that, while long-range triplet correlations are crucial, the mechanism that generates the triplets may not be relevant for formation of rim channels. Whether the long-range triplets are generated by magnetic non-collinearity between separate F layers (Nb/Ni/Cu/Co devices), the spin texture of a single ferromagnet (Nb/Co disks), or an intrinsic magnetic inhomogeneity at the S/F interface (LSMO-based disks), the rim currents appear as long as the long-range triplets are present.

\item[iii]
The size of the rim channels does not scale with the size of the disk; it appears to vary according to the material. As shown in Fig.~\ref{sqi-nbco}, the width of the rim channels ($\approx$~70 nm) is the same for Nb/Co devices, despite the difference in their diameters. However, when a 5 nm Cu layer is placed at the S/F interface, the channels become considerably wider (see Fig.~\ref{sqi-nbnico}).     

\item[iv]
We can induce supercurrents in halfmetallic LSMO disks, with very high current densities. The disk shows rim currents, similar to what we observe in Co disks. Puzzling is that, unlike in Co disks, the long-range triplet correlations are not suppressed when the disk is uniformly magnetized by a large in-plane field. Both the halfmetallicity and the oxide magnetism make this a different system, that invites further study.
\end{itemize}

\section{Acknowledgements}
This work was supported by the project `Spin texture Josephson junctions' (project number 680-91-128) and
by the Frontiers of Nanoscience (NanoFront) program, which are both (partly) financed by the Dutch Research Council (NWO). J. Y. is funded by the China Scholarship Council (No. 201808440424).The work was further supported by EU Cost actions CA16218 (NANOCOHYBRI) and CA21144 (SUPERQMAP). It benefitted from access to the Netherlands Centre for Electron
Nanoscopy (NeCEN) at Leiden University.

\end{document}